\documentclass[journal]{IEEEtran}

\usepackage{graphicx}
\usepackage{booktabs}
\usepackage{amsmath,amssymb}
\usepackage{fixltx2e}
\usepackage{float}
\usepackage{comment}
\usepackage{color} 
\usepackage{dblfloatfix}
\usepackage{float}
\usepackage{gensymb}
\usepackage{algorithm,algorithmic}
\usepackage{cite}

\newcommand{\bldW}{\mathbf W}
\newcommand{\bldDlt}{\boldsymbol \Delta}
\newcommand{\bldzero}{\boldsymbol 0}
\DeclareMathOperator{\var}{var}

\newcommand\ns[1] {{#1}}

\begin{document}

\title{Decentralized Picosecond Synchronization for Distributed Wireless Systems}
\author{Naim Shandi,~\IEEEmembership{Graduate~Student~Member,~IEEE,} Jason M.\ Merlo,~\IEEEmembership{Graduate~Student~Member,~IEEE,} \\and Jeffrey A.\ Nanzer,~\IEEEmembership{Senior Member,~IEEE}
\thanks{Manuscript received 2023.}
\thanks{This work was supported in part by Office of Naval Research under Grant \#N00014-20-1-2389, the National Science Foundation under Grant \#1751655, and Google under the Google Research Scholar program \textit{(Corresponding author: Jeffrey A. Nanzer)}}
	\thanks{	
	The authors are with the Department of Electrical and Computer Engineering, Michigan State University, East Lansing, MI 48824 USA (email: shandina@msu.edu, merlojas@msu.edu, nanzer@msu.edu).}
}

\maketitle

\begin{abstract}
We demonstrate a wireless, decentralized time-alignment method for distributed antenna arrays and distributed wireless networks that achieves picosecond-level synchronization. 
Distributed antenna arrays consist of spatially separated antennas that coordinate their functionality  at the wavelength level to achieve coherent operations such as distributed beamforming. 
Accurate time alignment (synchronization) of the local clocks on each node in the array is necessary to support accurate time-delay beamforming of modulated signals.
In this work we combine a consensus averaging algorithm and a high-accuracy wireless two-way time transfer method to achieve decentralized time alignment, correcting for the time-varying bias of the clocks in a method that has no central node. Internode time transfer is based on a spectrally-sparse, two-tone signal achieving near-optimal time delay accuracy. 
We experimentally demonstrate the approach in a wireless four-node software-defined radio system, with various network connectivity graphs. We show that within 20 iterations all the nodes achieve convergence within a bias of less than 12 ps and a standard deviation of less than 3 ps. The performance is evaluated versus the bandwidth of the two-tone waveform, which impacts the synchronization error, and versus the signal-to-noise ratio. 
\end{abstract}


\section{Introduction}

\begin{figure}[t!]
\centering
\includegraphics[width=1\columnwidth]{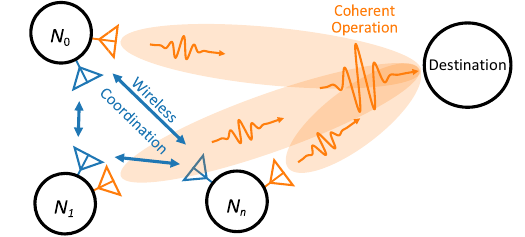}
\caption{\ns{Example of beamforming in coherent distributed antenna array system, showin an $n$-node distributed antenna array in open loop topology.} The nodes in the system perform wireless coordination to synchronize their time, frequency, and phase such that signals sum coherently at the destination in order to enable high-gain beamforming.}
\label{DAA}
\end{figure}

Recent interest in distributed antenna array systems has been based in part on the unique capabilities they offer for wireless applications. These applications include remote sensing, navigation, and global positioning \cite{nanzer2021distributed}. Distributed antenna arrays support increased signal gains at a lower cost in comparison with the monolithic array, increased failure tolerance, and the flexibility to dynamically reallocate the individual array elements \cite{nanzer2021distributed,ouassal2021decentralized}.  
A principal challenge in the design of distributed antenna array systems is the accurate synchronization of the array elements, enabling tightly coordinated operation. Coherent operation requires that the nodes (individual antennas) in the array should agree with respect to time, phase, and frequency; this can produce maximal power and information transfer at the destination of a target or receiving antenna (Fig.~\ref{DAA}). 
Two main coordination topologies have been developed in the past that differ in the way the network obtains feedback from the destination \cite{ouassal2021decentralized}. The closed loop topology uses explicit/implicit feedback from an active destination, while the open loop topology implements a self-alignment process without the need for feedback from the passive destination \cite{8378649,nanzer2017open}. Some closed-loop techniques include distributed transmitter synchronization, including receiver-coordinated explicit-feedback \cite{bidgare2012implementation}, one-bit feedback \cite{Mudumbai2006DistributedBU}, primary-secondary synchronization \cite{4202181}, reciprocity \cite{bidigare2015wideband}, round-trip synchronization \cite{4542555}, and two-way synchronization \cite{5957340}. However, they suffer from latency and overhead issues and moreover cannot be used for arbitrary beamsteering \cite{ouassal2021decentralized}. The open loop topology, on the other hand, operates securely with low synchronization overhead, enables arbitrary beamsteering, produces sparse antenna arrays, and lends itself to applications like remote sensing and radar where destination feedback is not available \cite{ouassal2021decentralized}.

Time synchronization in distributed phased arrays is a challenging problem, since appreciable alignment of the transmitted waveforms is necessary to achieve high gain, and for wideband signals this alignment translates to sub-nanosecond level synchronization~\cite{8378649}.
Beyond distributed phased arrays, time synchronization has received attention because of its role in other prominent applications such as wireless sensor networks~\cite{Simeone2008DistributedSI,wang-hu,estrin,mudumbai3}. Two classes of time synchronization are packet coupling time synchronization (local time is computed from timestamps carried on periodically exchanged packets) \cite{yener} and pulse-coupled synchronization \cite{hong, wang} which may employ phase-locked loops \cite{akaiwa3,06794,simeone3}. Applications of pulse-coupled time synchronization include change detection \cite{hong2} and data fusion \cite{murata}. Time transfer between systems can be accomplished in various ways~\cite{levine2008review}; in this work we consider two-way time synchronization because it implicitly compensates the channel parameters assuming the channel remains quasi-static during the time it takes for the synchronization to occur (i.e., the two-way message exchange is fast compared to the time over which the channel properties vary) ~\cite{merlo2022wireless}. In particular, the two-way time transfer approach is used in conjunction with a two-tone waveform {that achieves} near-optimal delay estimation accuracy~\cite{schlegel2019microwave}.
We furthermore consider microwave wireless systems~\cite{prager2020wireless,merlo2022wireless, Simeone2007DistributedTS, Simeone2008DistributedSI, 06794, wang,merlo2023picosecond, merlo2023wireless}, {which have neither the challenges associated with pointing, acquisition, and tracking, propagation effects, nor the size, weight, and cost limitations,} that are associated with optical approaches~\cite{giorgetta2013optical, shen, martha, kyle}. 
Typically, the development of methods for time synchronization has focused on centralized approaches, where nodes coordinate in a star or tree-based configuration. Such topologies, however, contain one or more single points of failure, which is undesirable for robust networking. 
To address this drawback, a decentralized approach is preferred, where nodes communicate locally with neighboring nodes to achieve group consensus without a central or primary node required~\cite{ouassal2021decentralized, 140489, 508788, wang}.

In this paper we present a new method for wireless time alignment of distributed antenna arrays. We use a consensus-averaging approach combined with the spectrally-sparse two-tone delay estimation waveform to produce a synchronization algorithm that coordinates nodes at the picosecond level in a fully decentralized topology. 
We provide a detailed description of the average consensus algorithm and the two-way time transfer approach, along with the time-delay estimation process using a two-tone waveform, and a refinement processe that serves to improve the delay estimation. 
The algorithm is implemented using software-defined radios (SDRs) and GNU Radio software to experimentally demonstrate coordination with under 12 ps of bias $+$ standard deviation and less than 3 ps of standard deviation. Multiple network topologies are considered in a four-node distributed array. Experiments were conducted using cabled connections and using wireless time transfer. Decentralized time-alignment method with maximum theoretical accuracy of the time delay estimation. 
The Cramer–Rao lower bound (CRLB), which provides the theoretical maximum accuracy of time delay estimation, serves to validate the empirical results and gauge the performance of the method. Finally, we evaluate the performance of the system as a function of the two-tone waveform bandwidth and the signal-to-noise ratio (SNR).

\section{Decentralized Time-Alignment Algorithm}

\subsection{System Model}

We model a distributed antenna array system as a graph $G=\{N,{\mathcal E} \}$ consisting of a set $N=\{1, 2, \ldots, n\}$ of $n$ nodes along with a set of edges {$\mathcal{E}$}. In this work we consider the edges to be undirected, meaning that for a pair of nodes $i$ and $j$, the edge $(i,j) \in {\mathcal E}$ indicates that the nodes are connected with bidirectional communication between them~\cite{ouassal2021decentralized}. \ns{Each node in the array has a local clock for which a linear model can be applied, approximating the relative clock drift of crystal oscillators over short periods of time. For any node $i$ in the array, the time offset can be represented by a function of the true global time $t$ \cite{merlo2022wireless} 
\begin{equation}\label{modeleq1}
T_i(t) = \alpha_i(t)t +\delta_i(t)+\nu_i(t)
\end{equation}
where $\alpha_i$ is the time-varying relative frequency scale, $\delta_i$ is the time-varying bias term, and $\nu_i(t)$ is a zero-mean noise term. The bias term $\delta_i(t)$ consists of initial system time offsets we wish to correct and a group of time-varying delays through the RF system (such as noise due to thermal, shot, flicker, plasma, and quantum effects~\cite{pozar2005microwave}).}

\ns{The time offset $\delta_i(t)$ has both a static component $\beta_i$ and a dynamic component $\zeta_i$. The dynamic component is due to factors including thermally generated time-varying internal delays, among others. The static components include constant system delays associated with the lengths of internal traces and cables; these can be corrected via calibration~\cite{merlo2022wireless}. We assume that $\delta_i(t)$ and any time-varying frequency offset are quasistatic over the synchronization epoch (see Sec.\ II-D below). In this work we focus on correcting the time offsets between the RF front-ends irrespective of the sources; to focus the work on time estimation explicitly, we assume that the nodes are syntonized (frequency locked), which may be accomplished wirelessly by various means ~\cite{mghabghab2021open-loop,alemdar2021rfclock,10443654,hassna2020decentralized}. The time at node $i$ can then be given by
\begin{equation}\label{modeleq2}
T_i(t) = t +\beta_i+\zeta_i(t)+\nu_i(t).
\end{equation}
Note that without correction for the dynamic component of the time-varying bias, the time offsets between different RF front-ends will drift over time even when the system is frequency syntonized and the static component is compensated for. Hence one significant aspect of the time synchronization method is the continuous compensation for the dynamic component of the time bias.}

\begin{figure}[t!]
\centering
\includegraphics[width=1.0\columnwidth]{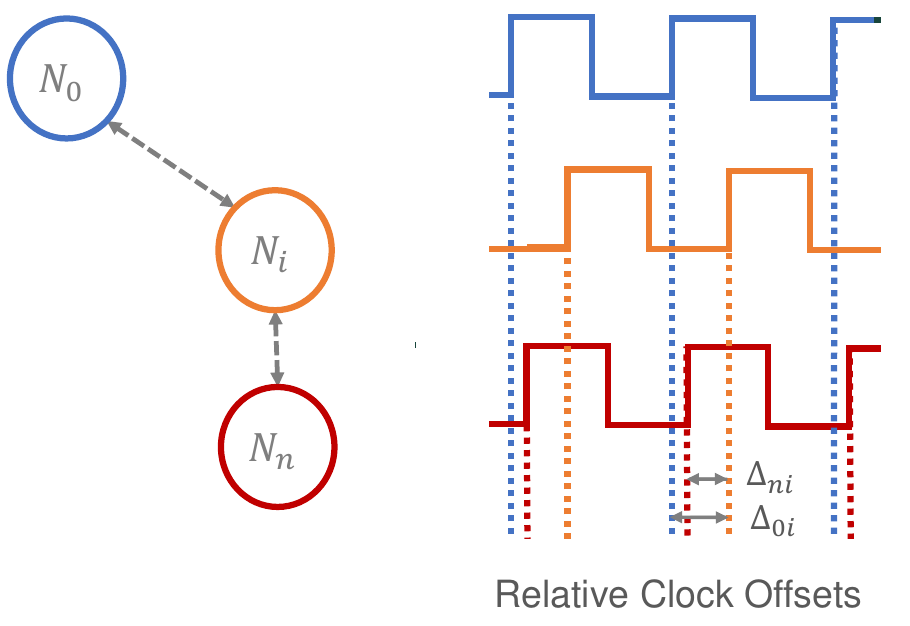}
\caption{The relative clock offsets at node $i$ in a distributed antenna array. The signed quantities $\Delta_{0i}$ and $\Delta_{ni}$ are the clock time offsets for node $i$ relative to nodes $0$ and $n$, respectively.}
\label{W_Delta}
\end{figure}

The relative time offset between node $i$ and any adjacent node $j$ in the neighborhood $\eta_i$ of node $i$ (i.e., all nodes that have a connection to node $i$) is denoted as $\Delta_{ji}, j\in\eta_i$. Fig.~\ref{W_Delta} shows the relative clock offsets between node $i$ and two adjacent nodes for a hypothetical array.

\subsection{Internode Offset Estimation}

\begin{figure}[t!]
\centering
\includegraphics[width=1.0\columnwidth]{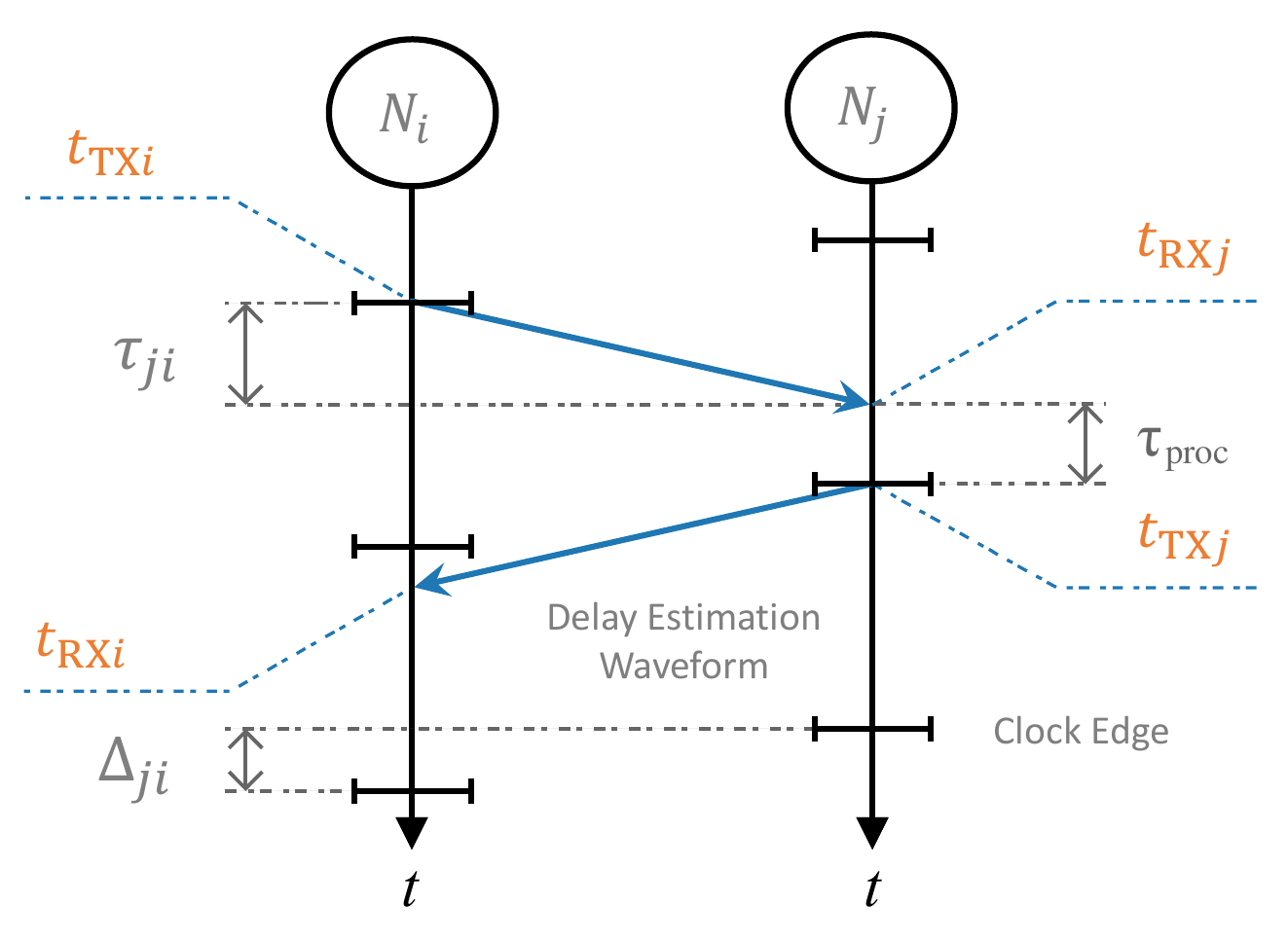}
\caption{Timing diagram describing the two-way time transfer method between node $i$ and node $j$ in a distributed antenna array. Node $i$ initiates time transfer with an adjacent node, $j$. A delay estimation waveform is transmitted from node $i$ to node $j$, which in turn transmits the delay waveform back to node $i$. Timestamps  $t_{{\rm TX}i}$, $t_{{\rm TX}j}$, $t_{{\rm RX}i}$, and $t_{{\rm RX}j}$, are saved at each transmission and reception; from these four timestamps, the relative time offset (as well as the time delay) can be computed.}
\label{timing_diagram}
\end{figure}

The {time} offset $\Delta_{ji}$ is determined by transferring clock time information between the nodes, which may be done in a one-way manner, where one node passes information to another, or in a two-way manner where the nodes exchange timing information. The two-way time transfer method was chosen in this work because the one-way time transfer method requires channel sounding to determine the characteristics of the propagation channel, and may also require precise knowledge of the relative positions and trajectories of the nodes in the system. In the two-way time transfer approach, the propagation path is reciprocal as long as the channel is quasistatic, and thus its impact cancels, as shown below; however, each node must be able to transmit and receive. Channel reciprocity is maintained as long as the platform motion is minimal relative to the beamforming wavelength during the synchronization epoch, which is a reasonable assumption for typical platform motion~\cite{levine2008review}. Fig.~\ref{timing_diagram} shows the two-way time transfer timing diagram. A delay estimation waveform is transmitted from node $i$ on a clock edge at time $t_{\text{TX}i}$, and is received by node $j$ at a time $t_{\text{RX}j}$. Node $j$ waits until a clock edge, and retransmits a signal at time $t_{\text{TX}j}$, which is received by node $i$ at time $t_{\text{RX}i}$. \ns{Because all the nodes in the network are controlled by one host computer, all four timestamps are processed in that computer and a network packet-based transmission method is not required for exchange of timestamps between the nodes.}

With the assumption that the channel is quasistatic during the synchronization epoch, the offsets between the two clocks can be obtained through a simple calculation using the four timestamps. The time offset between two nodes $i$ and $j$ is given by
\begin{equation}\label{offset}
 \Delta_{ji}=\frac{(t_{{\rm RX}j}-t_{{\rm TX}i})-(t_{{\rm RX}i}-t_{{\rm TX}j})}{2}
\end{equation}
After the estimation of the time offset between node $i$ {and} its adjacent connected nodes, the average consensus method is employed to compute the average of these time offsets and compensate for it by adding it to the local clock timing through an iterative process, as described below. Note that the time $\tau_{\text{proc}}$ required for a node to process a pulse is immaterial provided that the clock bias is quasistatic over the synchronization epoch.
The two-way time transfer method also provides a way to estimate the propagation delay (and thus the distance) between each pair of nodes in the array through the time delay estimation equation by simply averaging the apparent times of flight in each direction~\cite{merlo2022wireless}.

\subsection{Time Delay Estimation and Theoretical Limits on Accuracy}

As indicated by \eqref{offset}, the error associated with estimating the time offset between a node pair is determined by the error in estimating the two times $t_{{\rm RX}i}$, and $t_{{\rm RX}j}$, and thus the timing error is dependent on the accuracy of the delay estimator.
The theoretical performance limit for accurate estimation of time delays is given by the Cramer--Rao lower bound (CRLB) inequality
\begin{equation}\label{CRLB}
\var(\hat{\tau} -\tau) \ge \frac{N_0}{2 \zeta^2 E_s}
\end{equation}
where the left-hand side is the variance on the estimate of the delay. On the right-hand side $N_0$ is the noise power spectral density (PSD), $\zeta^2$ is the mean-squared bandwidth defined by the second moment of the signal spectrum, and $E_s$ is the signal energy. Essentially, the theoretical lower bound is dependent on the SNR and the waveform characteristics~\cite{nanzer-sharp}, \cite[Chapter 7.2]{Richards}. In the case of a matched filter, the ratio of the signal energy to the noise power spectral density becomes $E_s/N_0 = \text{SNR}$; substitution into \eqref{CRLB} yields
\begin{equation}\label{var}
\var(\hat{\tau} -\tau) \ge \frac{1}{2 \zeta^2 \text{SNR}},
\end{equation}
a relation which displays the inverse relationship between the product of the SNR and waveform mean-squared bandwidth, and the variance of the time delay estimation. 
Because the accuracy is inversely dependent on the mean-square bandwidth, which is the second moment of the signal spectrum, the accuracy is minimized (lowest error) for a given SNR when the second moment is maximized, which occurs when the signal energy is concentrated into two discrete frequencies with a wide frequency separation~\cite{nanzer-sharp}.
A two-tone signal closely approximates this ideal waveform in practical systems, where the bandwidth of each of the individual tones is inversely proportional to the signal duration. 
Besides minimizing time delay estimation error, the use of a two-tone waveform can reduce system hardware requirements when compared to systems which use conventional fully-occupied signal bandwidths such as linear frequency modulated (LFM) waveforms~\cite{schlegel2019microwave}. It can also supplement an existing waveform for joint communication and sensing applications due to the fact that the bandwidth between the two tones is left unoccupied~\cite{anton-william}. 

\ns{Performance bounds for time-delay estimation have received much attention. 
An extensive survey can be found in \cite{1706138}, which also  discusses how the classical estimation bounds can be modified in order to accommodate the coherence loss and reveal a threshold coherence phenomenon. Other papers present, for instance, a method for time-of-flight estimation bounds for wideband systems \cite{1164429} and a practical algorithm that approaches the Weinstein bounds \cite{6289099}.
The CRLB inequalities given in \eqref{CRLB} and \eqref{var} describe the lower bound of the time delay estimator for a single link in a network. For a fully connected multi-node network with $n$ nodes, the total number of links is $n(n-1)/2$. To evaluate the time delay estimation precision for the multi-node network studied here, a simple lower bound is derived from \eqref{var} based on averaging the variances of the links:
\begin{equation}\label{lb}
\var_{\text{ave}}(\hat{\tau} -\tau)\ge \sum_{d=1}^{d=D}\frac{1}{2 D \zeta^2 \text{SNR}_d},    
\end{equation}
where $D$ is the total number of connected links in the network. The lower bound in \eqref{lb} sets a theoretical lower limit on the time delay estimator for a multi-node network without considering the iterative process for the average consensus algorithm. 
The standard deviation of the estimated parameter at a single node is formulated in detail in~\cite[Chapter 7.2]{richards2014fundamentals}\cite{ouassal2021decentralized}, which consider a large number of samples as well as the effect of time reduction caused by the time-duplexing of connections through each link in the neighborhood.
The authors suggest that the number of samples can be reduced by increasing the average number of connections in the network. This formulation leads to increased error as the average number of connections increases. The current study evaluates the time delay estimation using a two-tone waveform for a fully connected multi-node network (i.e, all nodes have same number of connections) after achieving the steady state or convergence. For simplicity, and to provide a base to compare the performance of the system with the state of the art, \eqref{lb} is used to evaluate the precision of the studied system while accounting for the different link SNRs.}

To perform the time delay estimation, it is very important to estimate the time delay of the received signal (two-tone waveform) with a high level of accuracy. A discrete-time matched filter is used to maximize the signal energy at the sample most closely corresponding to the true time delay of the waveform. 
The sampling rate of the digitizer, however, limits the accuracy of the time delay estimation. The peak produced by the matched filter represents the time delay of the received signal with some error resulting from the sampling process, causing discretization errors that limit the accuracy at a certain lower bound regardless of increases in the SNR or mean-square bandwidth. Hardware and cost constraints prevent direct sampling at picosecond levels; \ns{thus, a two-stage processing approach is used to refine the discretized matched filter output. In the first stage a quadratic least-squares (QLS) interpolation fits a parabola through three sampling points consisting of the peak of the matched filter and the two adjacent points, with the peak of the QLS parabola, which is known in closed form, taken to represent the refined time delay estimate~\cite{moddemeijer1991sampled}~\cite[Chapter 7.2]{Richards} 
\begin{equation}\label{QLS}
\hat \tau =\frac{T_s}{2}\frac{s_{\text{MF}}[n_{\text{max}}-1]-s_{\text{MF}}[n_{\text{max}}+1]}{s_{\text{MF}}[n_{\text{max}}-1]-2s_{\text{MF}}[n_{\text{max}}]+s_{\text{MF}}[n_{\text{max}}+1]}.
\end{equation}
Here, $T_s$ is the sampling interval, $n_{\text{max}}=\text{argmax}\{s_{\text{MF}}[n]\}$, and 
$$
s_{\text{MF}}[n]=s_{\text{RX}}[n]\circledast s^*_{\text{TX}}[-n]
$$
where $s^*_{\text{TX}}[n]$ is the complex conjugate of the ideal transmitted signal and $s_{\text{RX}}[n]$ is the received waveform.
In the second stage, the residual bias created from the imperfect QLS representation of the matched filter output is corrected. The residual bias is fully determined by the waveform and sampling parameters, and since this is known a priori, it can be corrected for by using a precomputed lookup table~\cite{merlo2022wireless}.}

\subsection{Average Consensus Algorithm}

\ns{The main goal of this work is to implement a decentralized time alignment method that provides a high accuracy wireless time-alignment for a distributed antenna array system. Each node in a decentralized antenna array is required to estimate and correct for the average time offsets with respect to the adjacent nodes that are connected to it.} Here we describe the average consensus method of an $n$-node distributed antenna array.
The average consensus method expresses the average of the time offsets at node $i$ as $(\bldW\bldDlt)_{ii}$, where {$\bldW= [w_{ij}]$}  is the mixing matrix, which is an $n \times n$ real matrix with nonzero entries corresponding to the edges in the graph and self loops (diagonal entries). To ensure convergence, the mixing matrix must be symmetric ($\bldW=\bldW^T$), doubly stochastic (the elements of each row and column sum to unity, respectively), and decentralized in the sense that $w_{ij}=0$ if $i \ne j$ and nodes $i$ and $j$ are not connected~\cite{boyd,ouassal2021decentralized}. In this work, the mixing matrix was created through the use of the Metropolis--Hastings constant edge weight matrix to ensure that the matrix requires only local information~\cite{boyd}; it is represented as 
$$
w_{ji} = \begin{cases} 
\displaystyle\frac{1}{\max\{\deg(i),\deg(j)\}+1} & \text{if } (i,j) \in {\mathcal E}, \\[3pt]
0 & \text{if } (i,j) \notin {\mathcal E} \text{ and } i \ne j, \\
1-\displaystyle\sum_{j:j \ne i} w_{ji} & \text{if } i = j,
\end{cases} 
$$
where $\deg(i)$ is the number of edges connected to node $i$.

The decentralized time alignment algorithm is described in Algorithm \ref{dec_algorithm}. In this work it is assumed that the network graph is fixed such that the mixing matrix does not change {during an} iteration. The average time offset at node $i$ is described as the $i$th diagonal element of the $n \times n$ time offset matrix $\bldW \bldDlt$ which is estimated from all nodes at iteration $k$. 

\begin{algorithm}[t!]
\caption{Average Consensus Algorithm for Decentralized Time Alignment} 
\centering
\begin{algorithmic}
\STATE $k=0$
\WHILE{stopping criteria not satisfied}
\STATE $k=k+1$ 
\STATE $t_i(k) = t_i(k-1) + (\bldW \bldDlt)_{ii}(k-1)$
\ENDWHILE
\RETURN $t_i(k)$
\end{algorithmic}
\label{dec_algorithm}
\end{algorithm}

The consensus algorithm works in effect by calculating the average value of the offsets seen by a node in its neighborhood, after which the node adjusts its local time base by the average offset value. While the process is based only on local information sharing, as long as the graph is strongly connected (information can flow from a given node to any other node) and the mixing matrix satisfies the constraints indicated above, the network will converge to the global average in a static case. In realistic systems the time errors are dynamic, however if the synchronization interval is sufficiently shorter than the clock drift {they} can be assumed to be static. Furthermore, the drift may be alleviated by using a simultaneous syntonization (frequency alignment) approach.
The decentralized time alignment algorithm updates the time at each node during each iteration (synchronization epoch) until it reaches convergence. The nodes in the array reach consensus when the clock time in each node reaches the average of the initial clock times in all the nodes of the array or, in other words, when the time offsets between all nodes are relatively small ($\bldDlt \approx \bldzero$).

\subsection{Frequency Syntonization}

\ns{Coherent operation of a distributed array requires frequency syntonization (i.e., all local oscillators sharing the same frequency) to ensure that the transmitted signals sum at the destination with at least 90\% of the ideal gain or, equivalently, with less than $0.5$ dB degradation from the ideal case. Frequency syntonization is also needed for accurate estimation of the time delay of the received waveform used for time synchronization and ranging estimation by the matched filter. In general, the local oscillators have frequency offsets and experience slow stochastic frequency drift over time. Other work \cite{merlo2022wireless} has indicated that if the time synchronization is implemented with relatively fast periodicity and is sufficiently accurate to minimize the frequency drift, the distributed array can be directly synchronized by adjusting the oscillator phase. Wireless frequency syntonization may be accomplished by various methods such as using two-tone frequency locking~\cite{alemdar2021rfclock,mghabghab2021open-loop}, among others~\cite{10443654,hassna2020decentralized}. This work focuses on compensating for the time bias terms in \eqref{modeleq2}, using the two-way time transfer method. Therefore, in the clock model presented in \eqref{modeleq1}, the time-varying relative frequency $\alpha_i$ is kept constant because the oscillator frequency is locked using the centralized open-loop cabled frequency syntonization with a 10 MHz frequency reference.}

\section{Time Synchronization Experiments}

\subsection{Experimental Configuration}

The decentralized synchronization algorithm was evaluated experimentally using a four-node system. 
We evaluated the performance for a system with cabled connections, providing high SNR connections without environmental scattering, and also a wireless system. Various graph topologies were considered. 
The purpose of the cabled configuration is to validate the algorithm in a highly controlled environment without propagation aspects, while the wireless configuration better represented a distributed antenna array. The cabled experiment was implemented in a loop topology, with two connections per node, while the connection topology was varied for the wireless configuration. The experiment evaluates the number of iterations required to achieve convergence as well as the precision and accuracy of the method. 

\ns{We furthermore evaluated the performance as a function of the SNR and as a function of the two-tone waveform bandwidth, which impacts the delay estimation error. The SNR evaluation was implemented with an SNR (the average estimated SNR sweep of all the links in the system) from 14 dB to 36 dB using a 40 MHz two-tone waveform, and the two-tone bandwidth evaluation was implemented with the tone separation varied from 10 MHz to 50 MHz in 10 MHz increments with a fixed SNR of 36 dB. These results were compared to the theoretical error bounds given by \eqref{lb} and \eqref{var} with the same waveform and different link SNR values. Each of these experiments was repeated ten times, and the time offsets between the nodes of the system were recorded over 100 iterations. The mean standard deviation and the mean bias were calculated between the ten measurements. These parameters were computed from the estimated time offsets between the local clocks of the nodes in the system.}

\begin{figure}[t!]
\centering
\includegraphics[width=1.0\columnwidth]{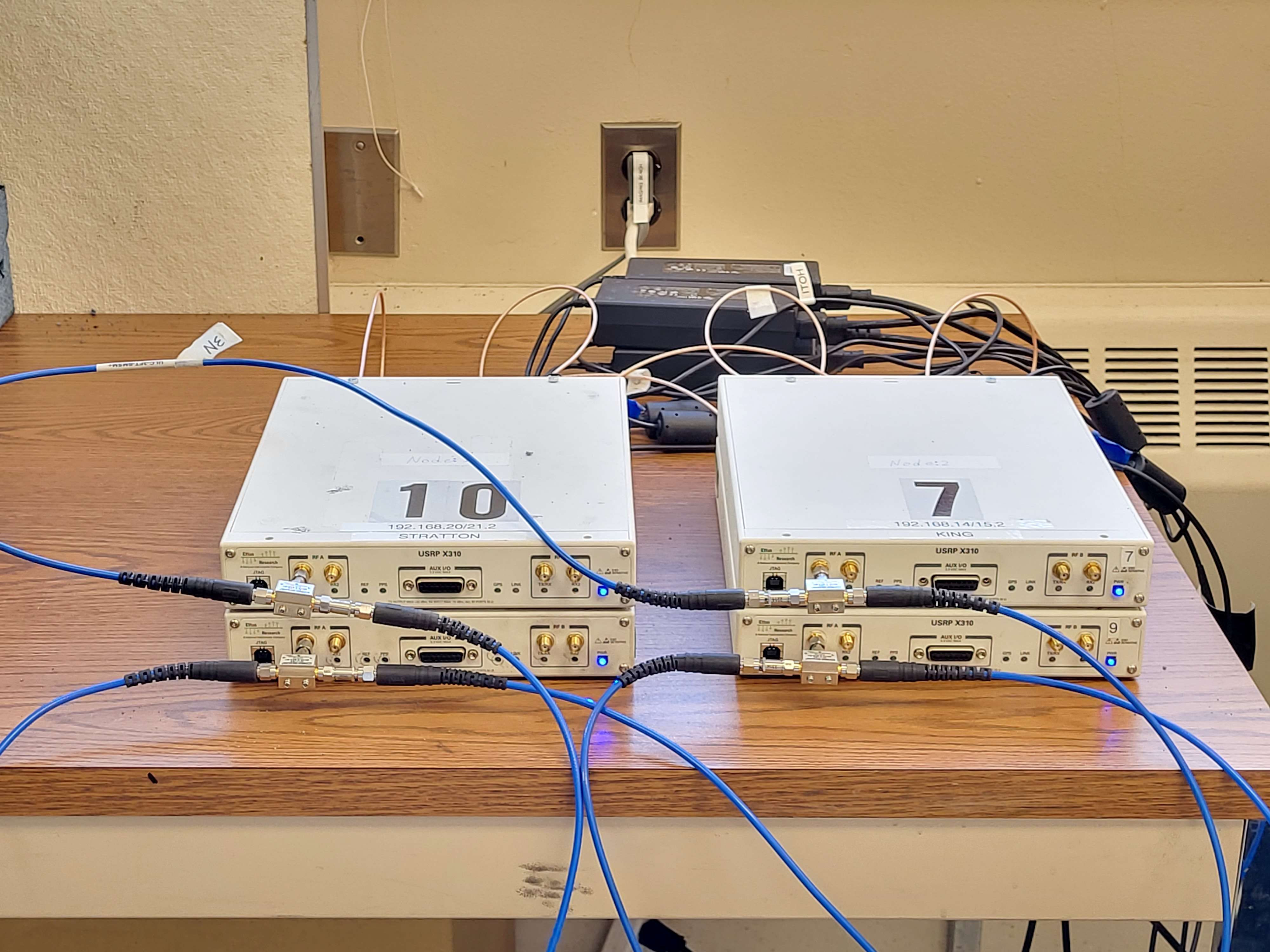}
\caption{Setup of the cabled time transfer experiment for the four-node distributed antenna array. The figure shows the four SDRs connected to four bidirectional two-to-one power splitters. The SDR TX/RX ports are protected by four 30 dB attenuators as shown in Fig.~ \ref{sc_wired_4c}}.
\label{wired_4c}
\end{figure}

\begin{figure}[t!]
\centering
\includegraphics[width=1.0\columnwidth]{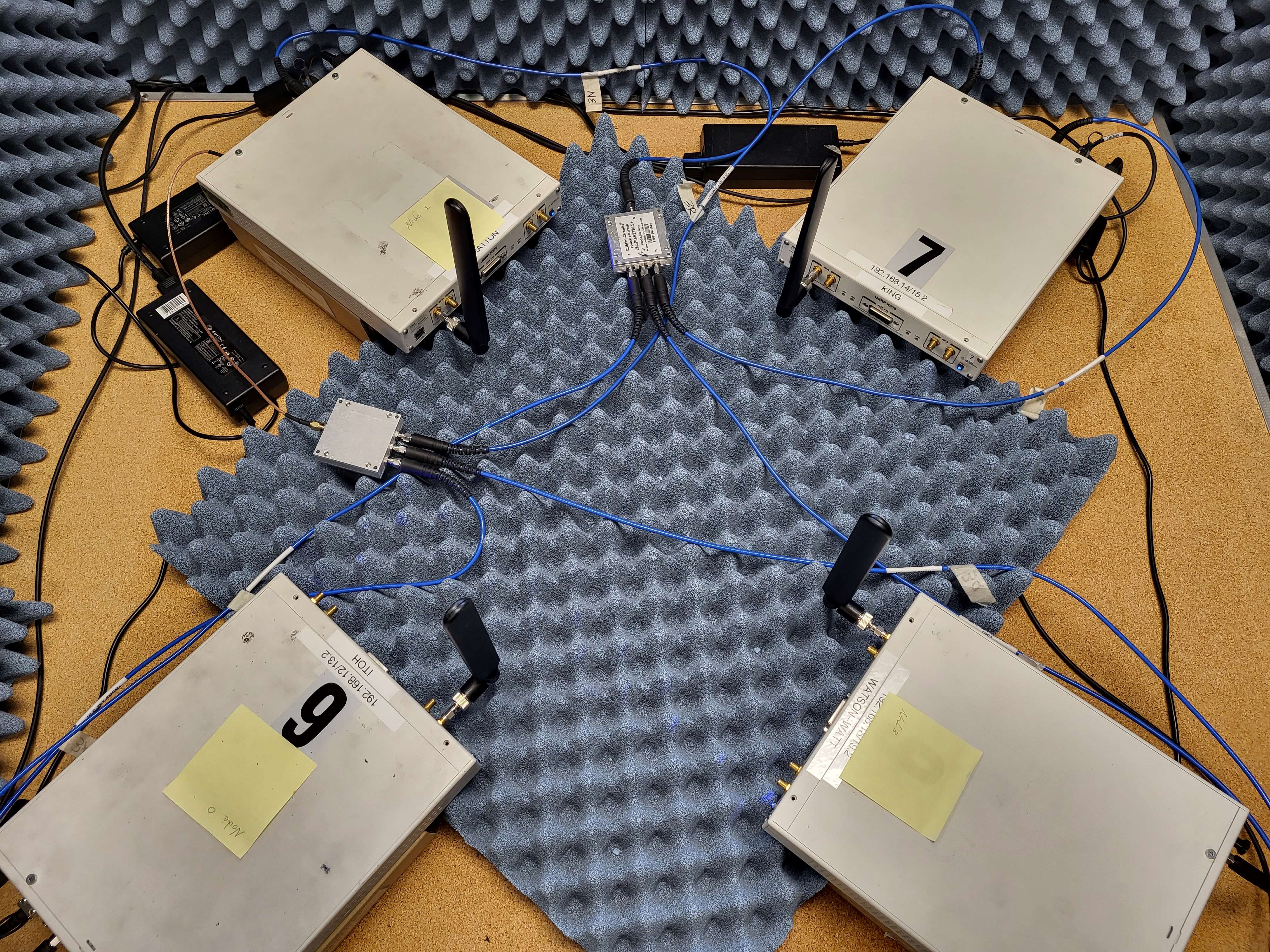}
\caption{Setup of the wireless time transfer experiment for four-node distributed antenna array. Four dipole antennas connected at the SDRs form TX/RX channels for the wireless two-way time transfer between the nodes. Two three-to-one power splitters are used for cabled frequency reference and cabled PPS for initial coarse time alignment.}
\label{wireless}
\end{figure}

\begin{figure}[t!]
\centering
\includegraphics[width=1.0\columnwidth]{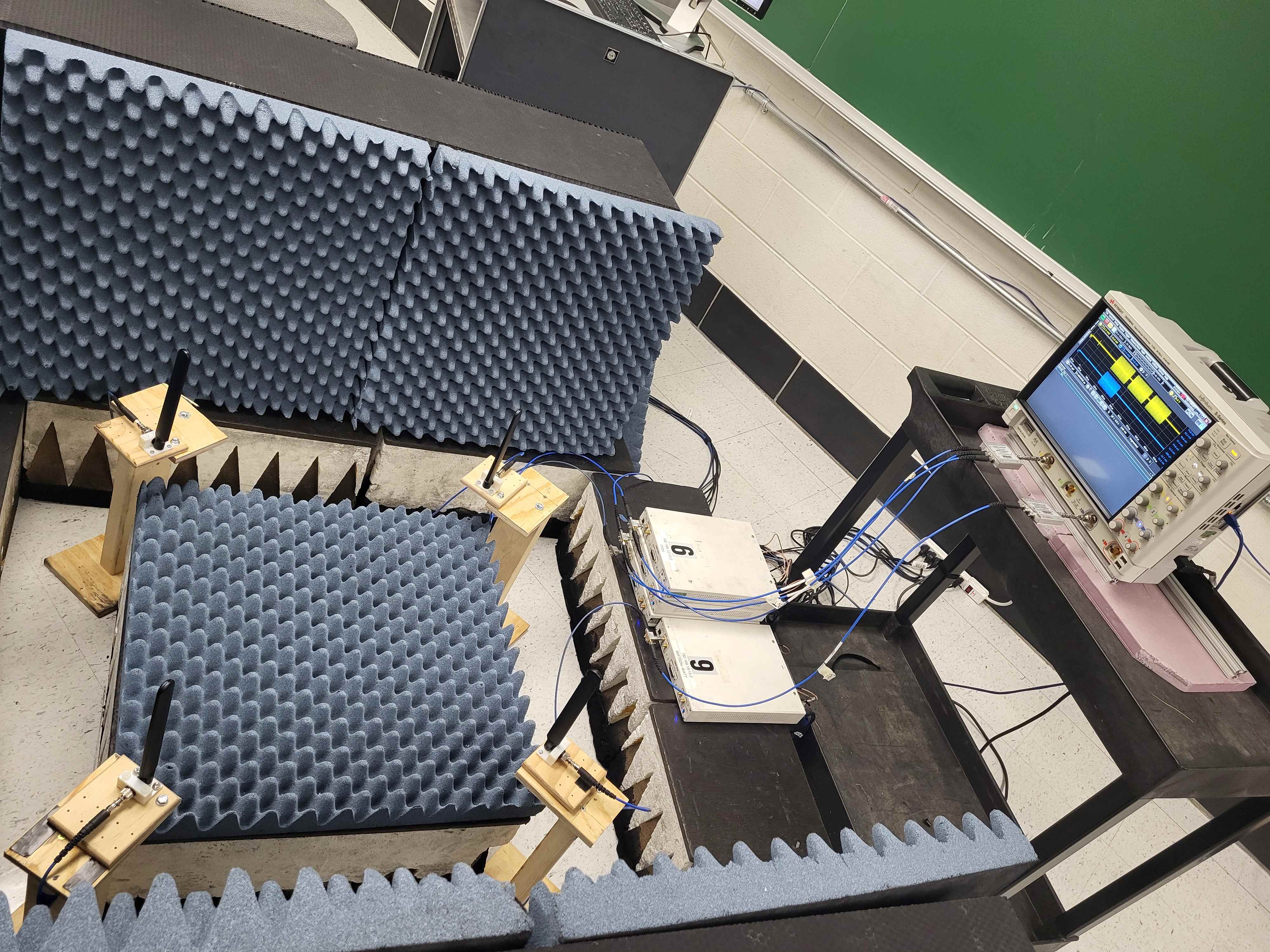}
\caption{\ns{Setup of the wireless time transfer experiment for four-node distributed antenna array with external verification. Verification waveforms were sent to a receiver (a Keysight DSOS8404A with sampling rate of 20 GSa/s) which is connected through cables to one end of the power splitter at the TX/RX port of each SDR. Nodes 0, 1, and 2 were connected to channel 1, and node 3 was connected to channel 3. Zero padding was applied to the verification signal to be aligned and saved separately within the oscilloscope window for post-processing purposes.}}
\label{verification}
\end{figure}

\begin{figure*}[t!]
\centering
\includegraphics[width=5in]{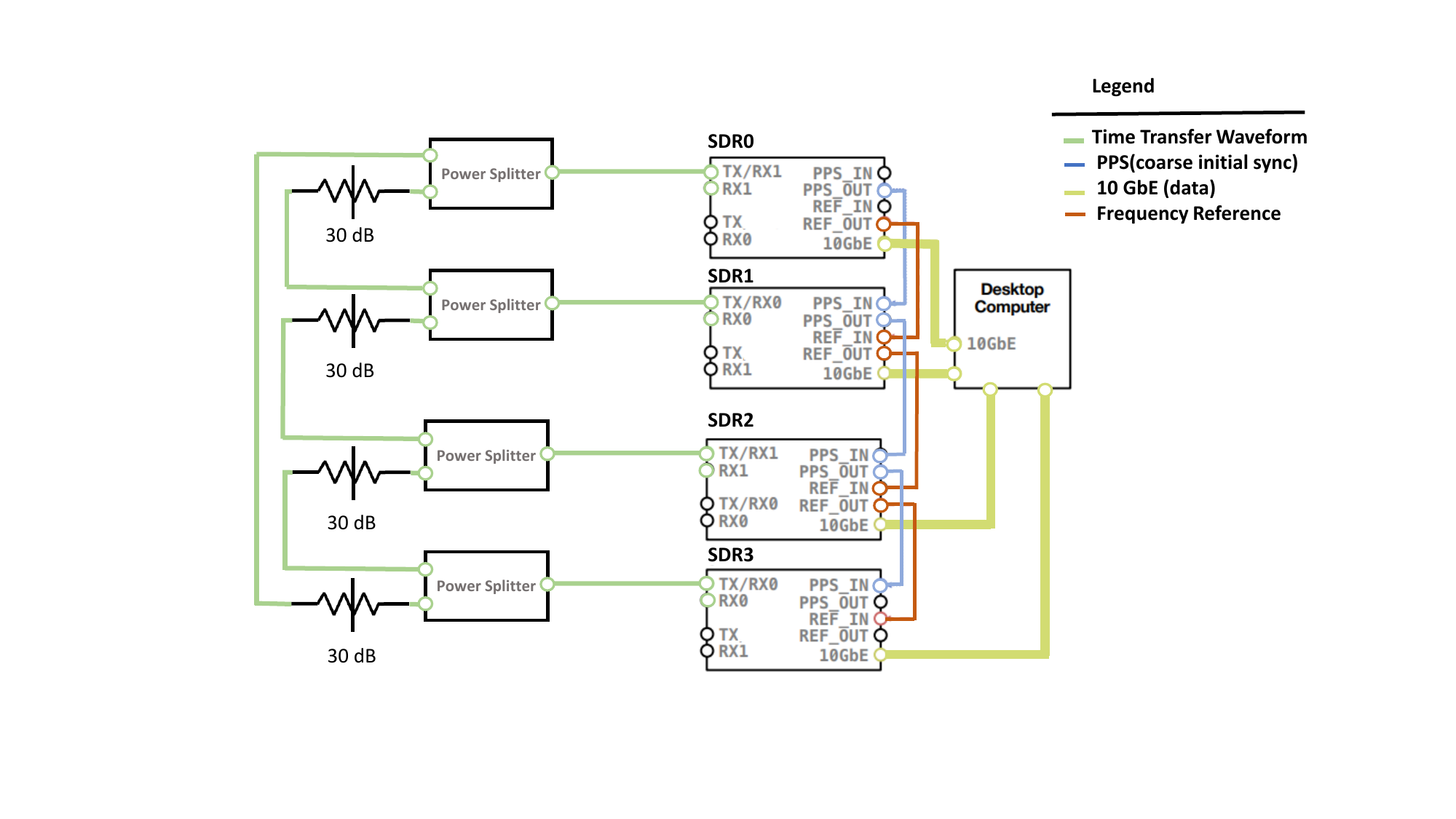}
\caption{Schematic of the cabled time transfer experiment for the four-node distributed antenna array. The figure shows the four SDRs interfaced with desktop computer (controlled by GNU Radio software). The SDRs have cabled frequency reference and cabled PPS for initial coarse time alignment.}
\label{sc_wired_4c}
\end{figure*}

\begin{figure*}[t!]
\centering
\includegraphics[width=5in]{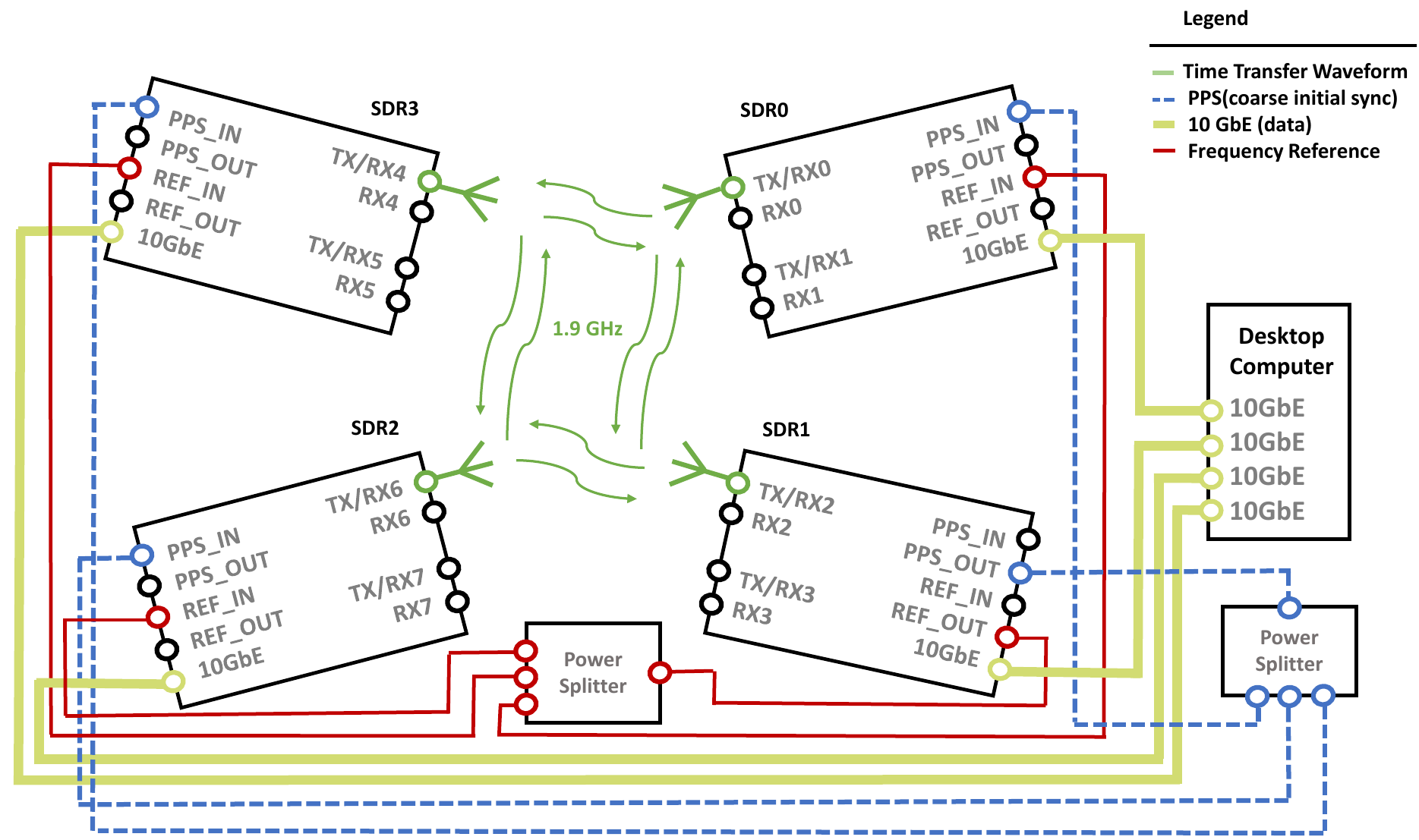}
\caption{Schematic of wireless time transfer experiment for four-node distributed antenna array. The four simulated SDRs are connected to dipole antennas for wireless time transfer at carrier frequency of $1.9$ GHz. The system has cabled frequency reference and cabled PPS for initial coarse time alignment.}
\label{sc_wireless}
\end{figure*}

\begin{figure*}[t!]
\centering
\includegraphics[width=5.5in]{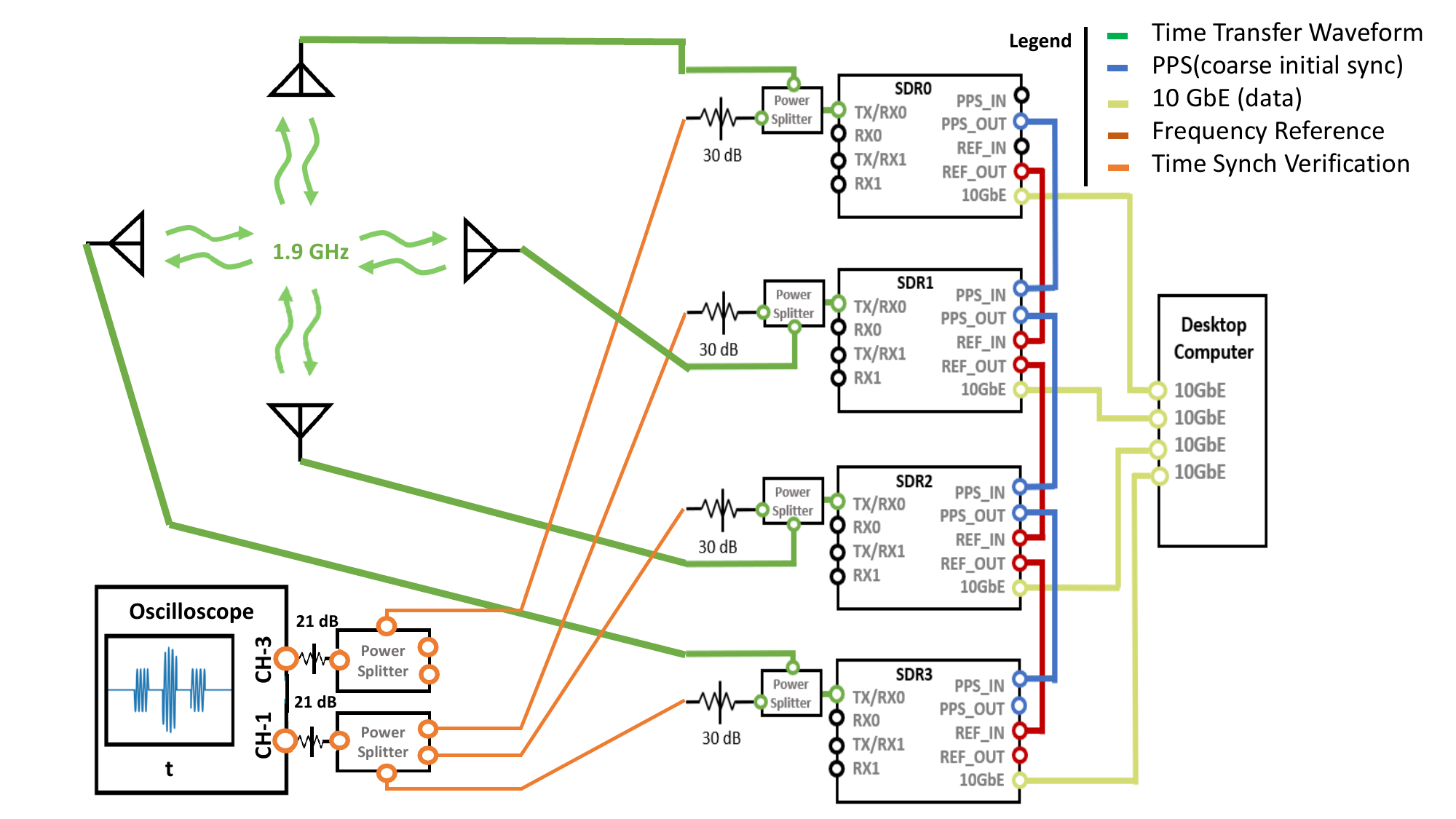}
\caption{\ns{Schematic of wireless time transfer experiment for four-node distributed antenna array. The four simulated SDRs are connected to dipole antennas for wireless time transfer at carrier frequency $1.9$ GHz. The system has cabled frequency reference and cabled PPS for initial coarse time alignment. An oscilloscope was used to verify the clock time synchronization externally. Verification pulses were sent by all the nodes in the array to channels 1 and 3 on the oscilloscope through cables with zero-padding such that they arrived within the 60 $\mu$s oscilloscope window. Verification pulses are saved and post processed to estimate the clock offsets between the nodes in the array.}}
\label{sc_verification}
\end{figure*}

Syntonization of the systems was implemented using cabling in all the configurations as shown in the experimental setups and schematics in Figs.~\ref{wired_4c}--\ref{sc_verification}, which alleviated clock drift, but not the time biases, since the focus of this is on synchronizing the clocks. Wireless syntonization can be implemented using various techniques (e.g.,~\cite{9670698}).

\begin{figure*}[t!]
\centering
\includegraphics[width=2.0\columnwidth]{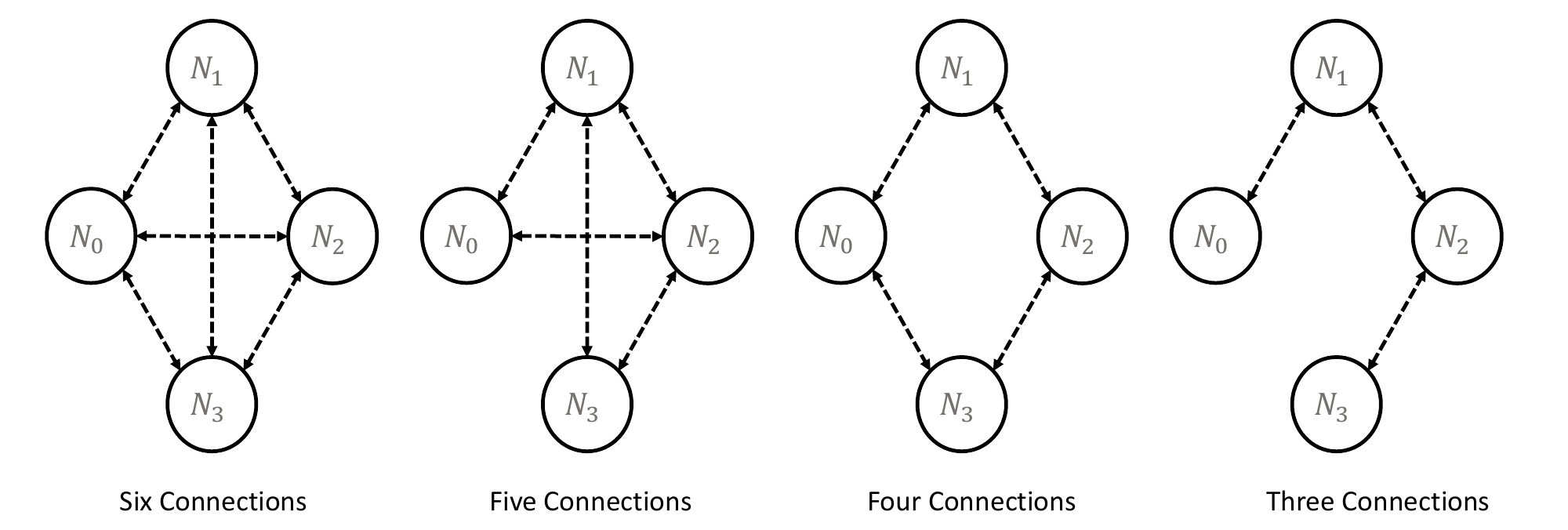}
\caption{The four connectivity configurations tested in the wireless experiments.}
\label{conn}
\end{figure*}

The experiments represented a four-node distributed antenna array and employed four SDRs (Ettus Research Universal Software Radio Peripheral (USRP) X310). Each SDR had two UBX-160 daughterboards supporting 160 MHz of instantaneous analog bandwidth. The sampling rate of the X310 used for this experiment was 200 MSa/s. For the cabled time-transfer experiment, power splitters were used to connect each of the nodes to the adjacent nodes as required to achieve the four connections described in Fig.~\ref{conn}. In addition to the power splitters, 30-dB attenuators were used to protect the SDRs during the cabled time transfer experiments. The four SDRs were connected through a cable for a pulse-per-second (PPS) coarse initial time alignment (see Figs.~\ref{sc_wired_4c} and \ref{sc_wireless}). The PPS cable was used once at the start of the experiment to align the local clocks of the system to within {a} few clock ticks. The initial coarse alignment is required for time synchronization to ensure that the transmitted synchronization pulses arrive within the receive window. The accuracy of the PPS initial alignment is on the order of tens of nanoseconds, which is many orders of magnitude greater than that needed for coherent beamforming at microwave frequencies~\cite{8378649}. Initial time alignment can also be achieved using a method such as GNSS PPS synchronization or adjunct ultrawideband (UWB) transmitters, among other techniques. Four multiband swivel mount dipole antennas (SPDA24700/2700) were used during the wireless time transfer experiment. \ns{The time synchronization method was verified externally using a Keysight DSOS8404A with a sampling rate
of 20 GSa/s as a receiver (Figs.\ \ref{verification} and \ref{sc_verification}). The TX/RX RF channel of each SDR was
connected to a one-to-two power splitter with one output connected to a dipole antenna and the other connected to the oscilloscope channel. Nodes 0, 1, and 2 were connected to channel 1 of the oscilloscope through a three-to-one power splitter,
and node 3 was connected to channel 3. The SDRs sent 10 $\mu$s two-tone waveforms with zero padding to align the verification pulses within the 60 $\mu$s window. This alignment was necessary to save and post process the received pulses and to estimate the clock offsets between the nodes in the array. The external verification method demonstrated the improvement in system performance with the proposed time synchronization method by comparing it with the performance of the system without continuous time synchronization.}

\begin{figure}[t!]
\centering
\includegraphics[width=1.0\columnwidth]{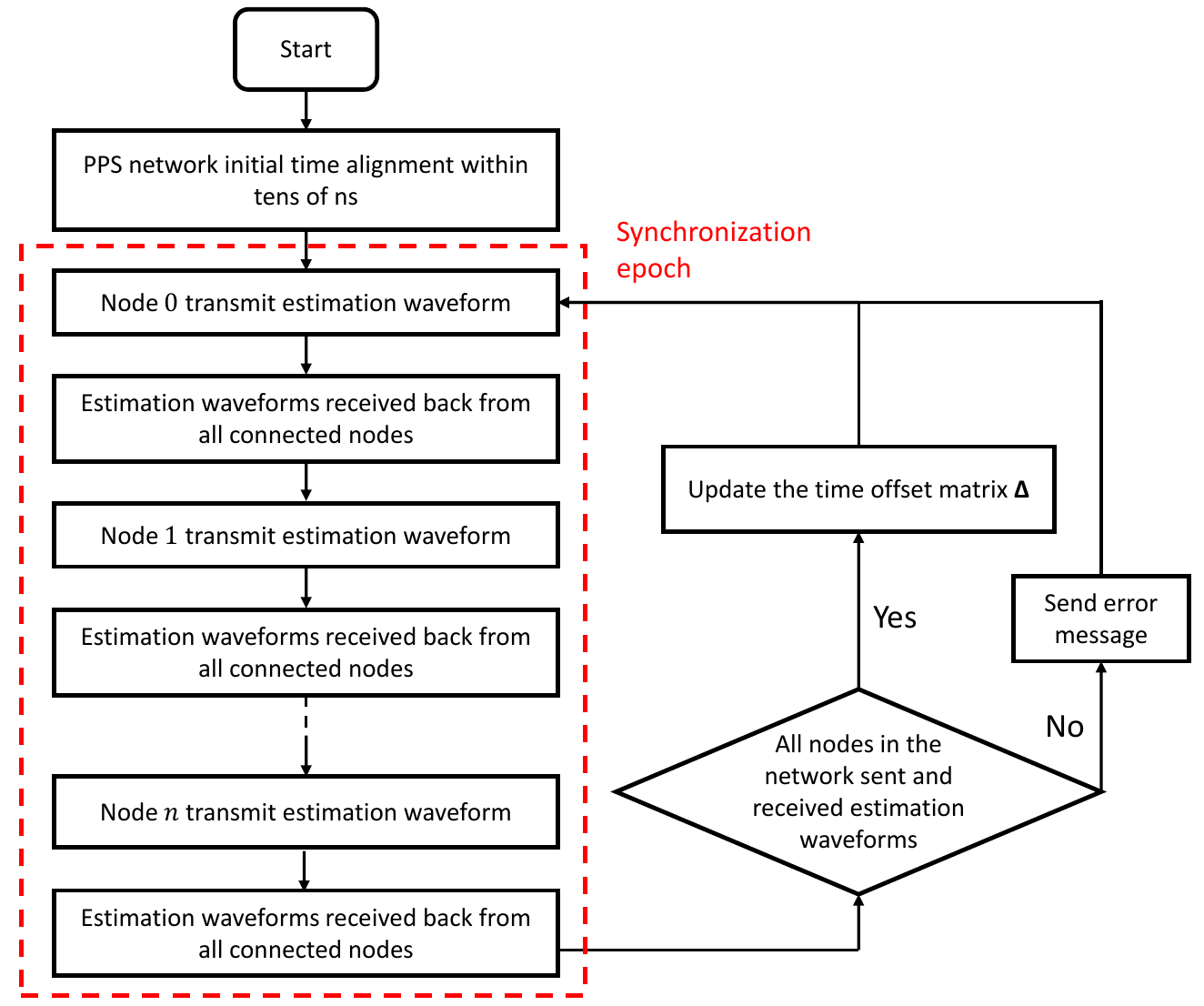}
\caption{\ns{Flowchart describing the implementation of the half-duplex scheme (TDMA), the synchronization epoch, and the process of updating the time offsets $\Delta_{ji}$}.}
\label{TDMA}
\end{figure}

The SDRs were connected through 10-Gbit Ethernet cable to a desktop computer for control and data processing. The computer ran Ubuntu 20.04 as its operating system. To process the data in real time, GNU Radio 3.10 software was interfaced with the Ettus Universal Software Radio Peripheral (USRP) Hardware Driver (UHD) 4.6.0.0 of the SDRs. The SDRs were programmed for bursty operation, using scheduled timed transmissions and receptions to achieve the full capability of the SDRs sampling rate of 200 MSa/s in order to maximize timing accuracy. \ns{A half-duplex scheme was employed using the time-division multiple access (TDMA) method. TDMA was implemented to schedule signal transmission in the network such that only one SDR could transmit at a time (Fig.\ \ref{TDMA}). Using the TDMA method, each node is identified by its time slot; thus no additional information had to be  modulated onto the time transfer waveform to identify each node. 
The initial acquisition component of the TDMA process was implemented by using the PPS inputs cabled between devices. Fig.\ \ref{TDMA} also shows the process of updating the local time offsets $\Delta_{ji}$, which were stored in the control computer and added to the scheduled transmit times for accurate clock alignment in the software. This manual update was necessary due to the limitation in the API which did not support the in-place clock updates \cite{merlo2022wireless}.}

\ns{The baseband two-tone waveform was generated using GNU Radio software by combining two complex sinusoidal signals:
$$
S_\text{Two-Tone} = \tfrac{1}{2} [ e^{j(-2 \pi f t + \phi)} + e^{j(2 \pi f t + \phi)} ]
$$
where $f$ is one-half the bandwidth of the ranging waveform and $\phi$ is the phase angle. A single pulse was used to estimate each time delay during every iteration, with 5 ns rise and fall times chosen for the envelope of the transferred waveform in order to approximate a matched filtering effect with devices having finite switching times. Moreover, the time of arrival between sample bins can be estimated more accurately by extending the rising edge of the envelope across multiple samples in the presence of amplitude modulation\cite{merlo2022wireless}.}

A similar software implementation to the two-way time transfer process in \cite{merlo2022wireless} was implemented in this work. Due to the API limitations a local clock update process was not supported, and the local time offsets were stored in software and manually added to the scheduled transmit times on the control computer to align the clocks in the software.

The two-way time transfer method described in Section II was employed using a two-tone waveform. The pulsed two-tone signal parameters are listed in Table~\ref{tablewaveformparams}. For the first two experiments, the tone separation was fixed at 40 MHz and the SNR was set to 33 dB $\pm$ 3 dB for cabled configuration and 32 dB $\pm$ 4 dB for the wireless configuration. \ns{To evaluate the system performance, two statistical analyses were conducted on the measured data. First, the time transfer method precision was evaluated by averaging the standard deviations of the means of the time offsets between the nodes from ten repeated measurements over 100 iterations. Second, the accuracy of the method was evaluated by computing the average of the standard deviations and adding this to the error of the time offsets between the four nodes.}

\begin{table}[t!]
\caption{Waveform Parameters}
\begin{center}
\begin{tabular}{ll}
\toprule
Parameter & Value\\ \midrule
Waveform Type & Pulsed Two-Tone \\
Carrier Frequency & 1.9 GHz \\
Pulse Duration & 10 $\mu$s \\
Rise and Fall Times & 5 ns \\
Resynchronization Period & 500 ms \\
RX and TX Sample Rates$^*$ & 200 MSa/s \\
\bottomrule 
\end{tabular}
\end{center}
\par\vskip2pt
\hskip3.5pc \parbox{2.3in}{$^*$TX Digitally upsampled from 200 to 800 MSa/s \\ \phantom{$^*$}on device}
\label{tablewaveformparams}
\end{table}

\subsection{Experimental Results}

\begin{figure}[t!]
\centering
\includegraphics[width=1.0\columnwidth]{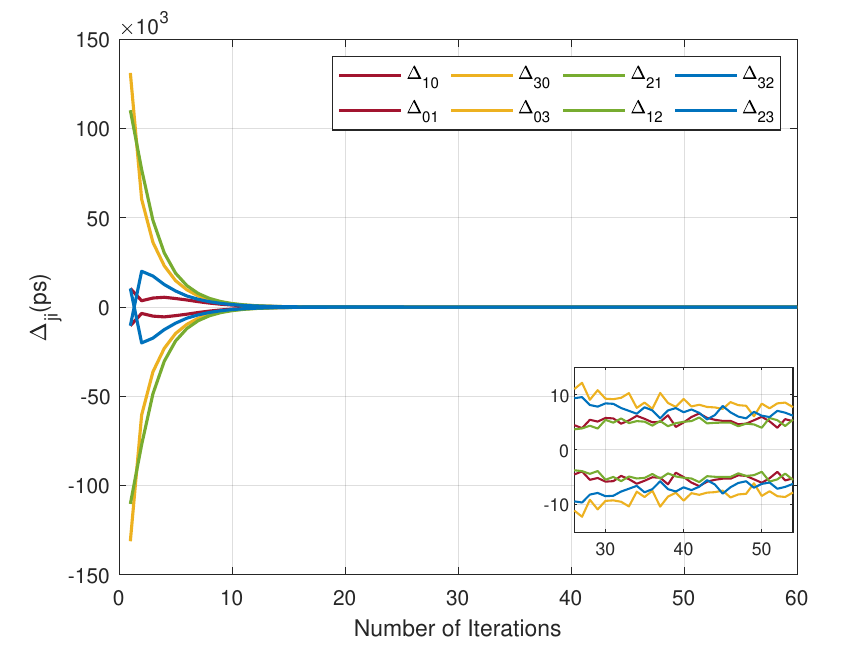}
\caption{The time offsets $\Delta_{ji}$ between the connected nodes of a four-node distributed antenna array over 60 iterations for cabled time transfer at 36 dB SNR and 40 MHz tone separation.}
\label{delta_cabled}
\end{figure}

\begin{figure}[t!]
\centering
\includegraphics[width=1.0\columnwidth]{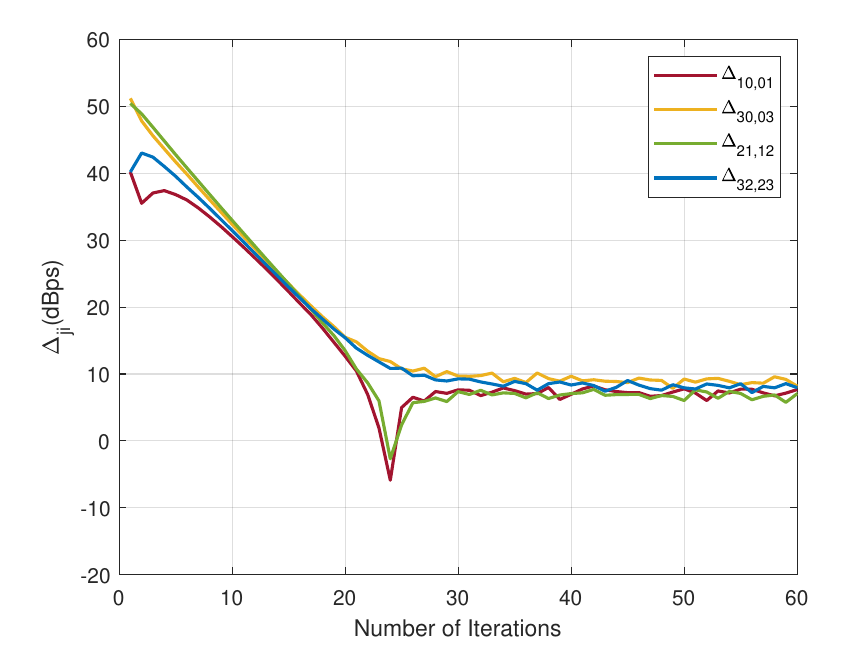}
\caption{The log magnitude of the time offsets $\Delta_{ji}$ between the connected nodes of a four-node distributed antenna array over 60 iterations for cabled time transfer at 36 dB SNR and 40 MHz tone separation.}
\label{delta_cabled_dB}
\end{figure}

\begin{figure}[t!]
\centering
\includegraphics[width=1.0\columnwidth]{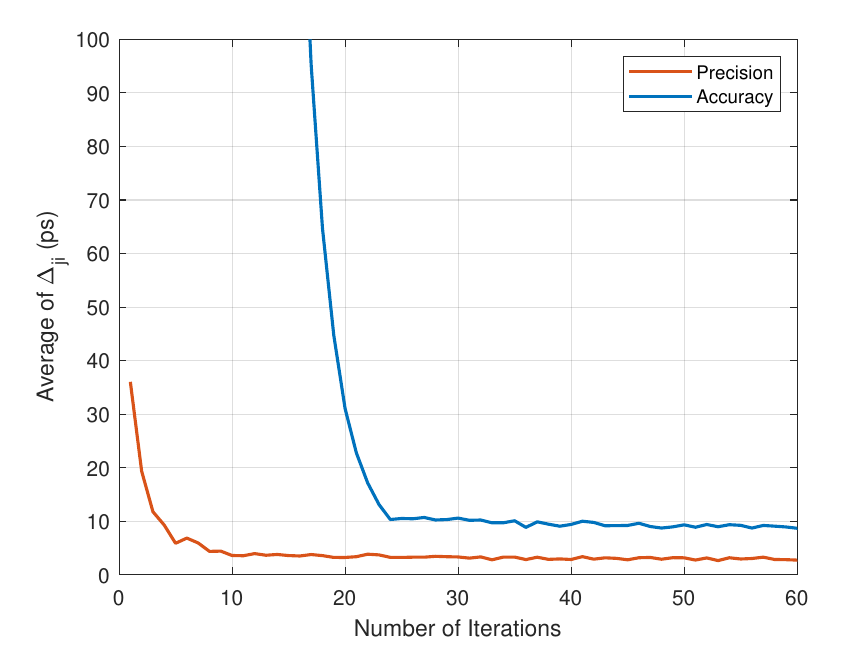}
\caption{The average accuracy (bias $+$ standard deviation) and precision (standard deviation) of the time offsets $\Delta_{ji}$ between the connected nodes of a four-node distributed antenna array over 60 iterations for cabled time transfer at 36 dB SNR and 40 MHz tone separation.}
\label{delta_cabled_ap}
\end{figure}

\ns{Each of the experiments was repeated ten times with a synchronization epoch of approximately 200 ms, and the time offsets between the nodes were measured over less than 2 minutes (60 iterations for the first two configurations, 100 iterations for the third and fourth configurations). The time delay estimation waveform parameters are listed in Table~\ref{tablewaveformparams}. The measured time offsets (between the connected nodes only) from the cabled time transfer experiment are plotted over 60 iterations in Fig.~\ref{delta_cabled}, where the time differences between the nodes converge around zero, to the picosecond level, as expected. The inset exhibits a magnified section of the same plot after 30 iterations, and shows that the time offsets among the four nodes
converge in about 32 iterations with error of less than 10 ps. Additional clarity regarding the convergence behavior is provided in Fig~\ref{delta_cabled_dB}, where the log magnitudes of the time offsets between the connected nodes are plotted over 60 iterations in dBps units. This figure confirms that the time differences between the nodes converge after about 32 iterations with errors of less than 10 ps. The precision and accuracy for cabled time transfer were computed and plotted over the same number of iterations by computing the average of the standard deviations and the average of the standard deviations plus the bias for precision and accuracy, respectively (see Fig.~\ref{delta_cabled_ap}). Note that the time offsets evaluated in these experiments are those measured internally by the system and are the compensated values.}

\begin{figure}[t!]
\centering
\includegraphics[width=1.0\columnwidth]{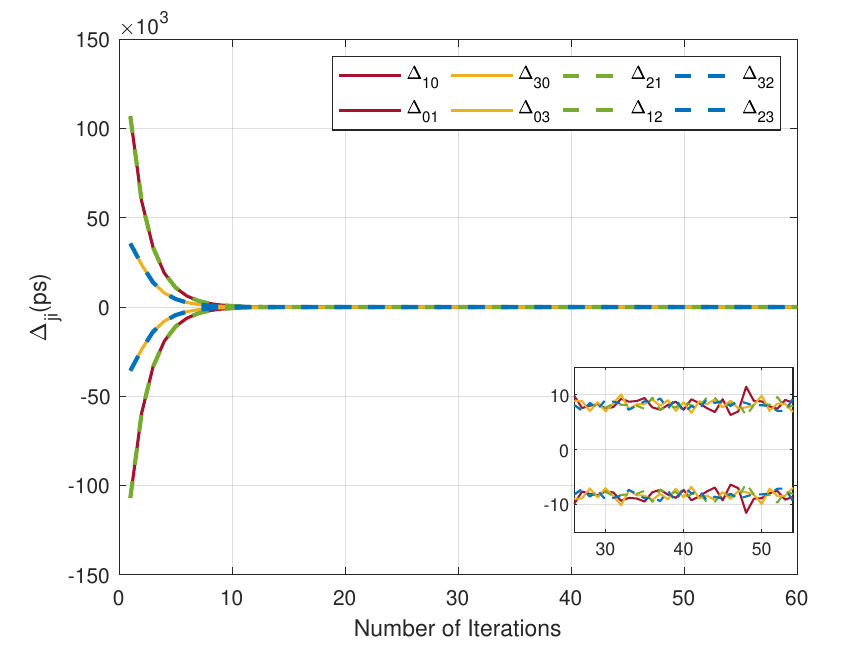}
\caption{The time offsets $\Delta_{ji}$ between the connected nodes of a four-node distributed antenna array over 60 iterations for wireless time transfer at 36 dB SNR and 40 MHz tone separation.}
\label{delta_wirless}
\end{figure}

\begin{figure}[t!]
\centering
\includegraphics[width=1.0\columnwidth]{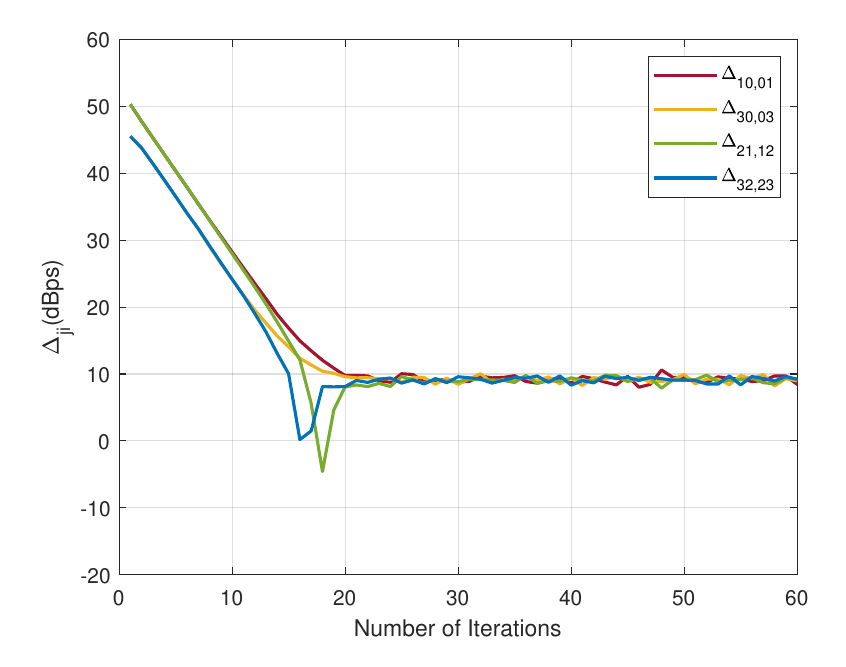}
\caption{The log magnitude of the time offsets $\Delta_{ji}$ between the connected nodes of a four-node distributed antenna array over 60 iterations for wireless time transfer at 36 dB SNR and 40 MHz tone separation.}
\label{delta_wireless_dB}
\end{figure}

\begin{figure}[t!]
\centering
\includegraphics[width=1.0\columnwidth]{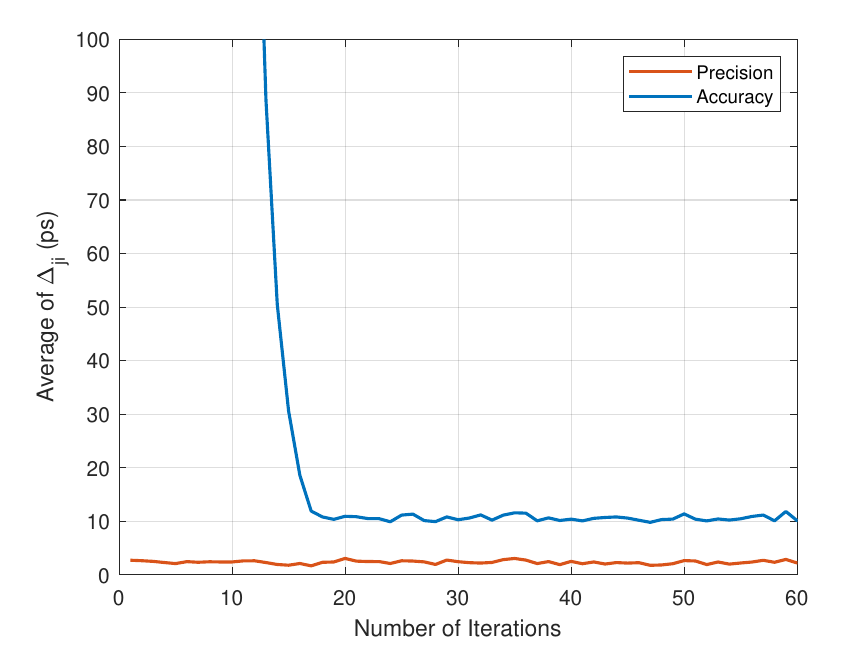}
\caption{The average accuracy (bias $+$ standard deviation) and precision (standard deviation) of the time offsets $\Delta_{ji}$ between the connected nodes of a four-node distributed antenna array over 60 iterations for wireless time transfer at 36 dB SNR and 40 MHz tone separation.}
\label{delta_wireless_ap}
\end{figure}

Figs.~\ref{delta_wirless}--\ref{delta_wireless_ap} show the results of the same measurements and computations repeated with the implementation of wireless configuration using dipole antennas.  Similar convergence behavior was noted after about 20 iterations with error of less than 12 ps.  

\begin{figure}[t!]
\centering
\includegraphics[width=1.0\columnwidth]{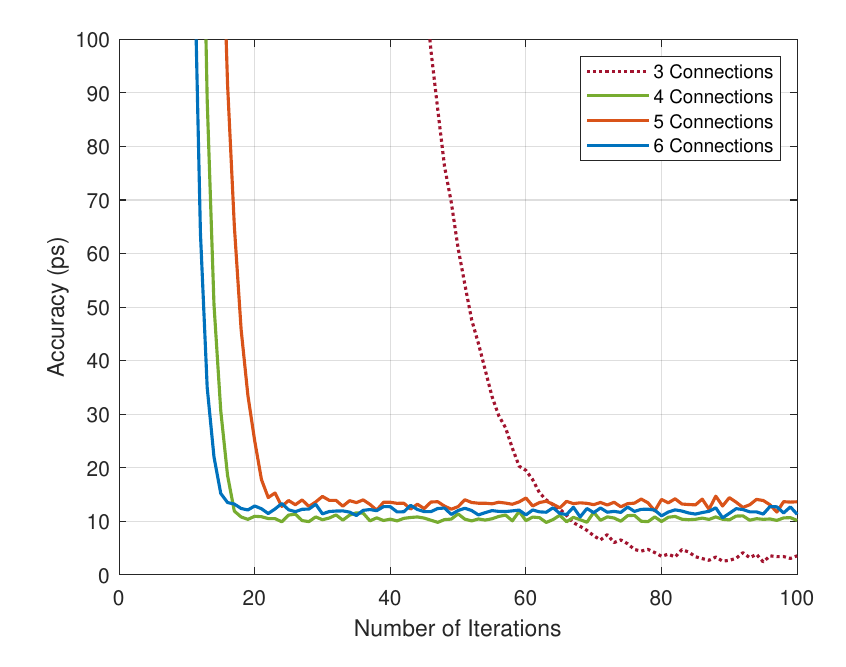}
\caption{Total residual error of the four possible connectivities among the nodes of a four-node distributed antenna array. The plot shows the average standard deviations plus error between the time offsets obtained from ten measurements for wireless time transfer configurations over 100 iterations.}
\label{diff_conn_a}
\end{figure}

\begin{figure}[t!]
\centering
\includegraphics[width=1.0\columnwidth]{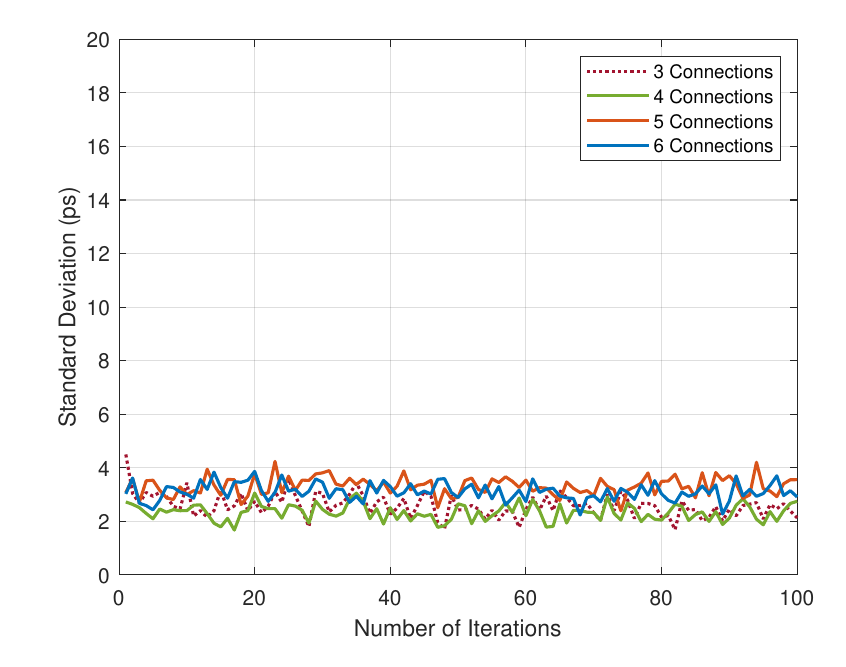}
\caption{Standard deviation of the four possible connectivities among the nodes of a four-node distributed antenna array. The plot shows the average standard deviations between the time offsets obtained from ten measurements for wireless time transfer configurations over 100 iterations.}
\label{diff_conn_p}
\end{figure}

The wireless time transfer experiment was repeated while changing the connectivity between the four nodes of the distributed antenna array. The connectivity graphs that were simulated are depicted in Fig.~\ref{conn}. Connections between nodes were either used or ignored in software to generate the desired topology. The accuracy comparison between the cases of three, four, five, and six connections is presented in Fig.~\ref{diff_conn_a}. This figure shows the effect of increasing the connectivity between the nodes on the number of iterations required to achieve the convergence as well as the accuracy of the results. It is worth noting that as the connectivity increases, the number of iterations required to achieve convergence decreases and the error increases (as the latency and uncertainty increase with the number of channels) in general. When a small number of nodes is used as in this experiment, the effect of connectivity becomes less significant as the connectivity increases insignificantly as seen in Fig.~\ref{diff_conn_a}. This explains why the four-connection configuration reached convergence with less error and fewer iterations compared with the five-connection configuration. In Fig.~\ref{diff_conn_p} the comparison of the average of the standard deviations over 100 iterations is plotted and shows less than 5 ps error obtained from varying the connectivity for the same evaluated four-node distributed array. Lastly note that for the case of six connections (a fully connected network), on theoretical grounds convergence may be expected after one iteration for ideal weighting. However, Algorithm 1 computes the average of the time differences between the node including the fact that $\Delta_{ii} = 0$ for each $i$ (the diagonals are zero) while the mixing matrix \textbf{W} is an all-ones matrix, which maintains an even weighting but slows the convergence time. This combines with system latency in the physical system to produce a slower than expected convergence rate for a fully connected network. Nonetheless, the network converges and maintains convergence with low timing error.

\begin{figure}[t!]
\centering
\includegraphics[width=1.0\columnwidth]{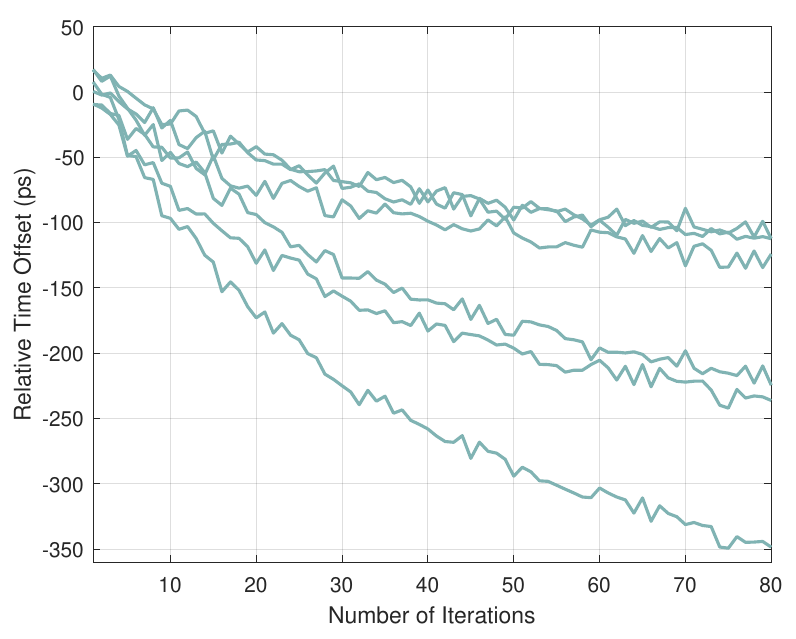}
\caption{\ns{The clock offsets between the four node of the system, measured and evaluated externally using an oscilloscope. The system is frequency syntonized and time synchronized only at the start of the synchronization process, after which the time synchronization algorithm is disabled.}}
\label{without_time_sync}
\end{figure}

\begin{figure}[t!]
\centering
\includegraphics[width=1.0\columnwidth]{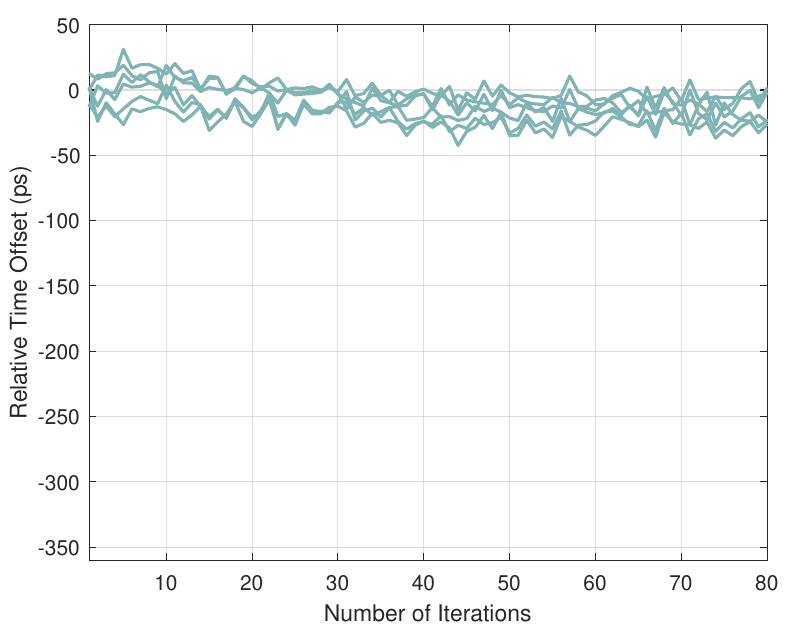}
\caption{\ns{The clock offsets between the four node of the system, measured and evaluated externally using an oscilloscope. This system was frequency locked and continuously time synchronized (the system continuously measured and corrected for the dynamic component of the time bias).}}
\label{with_time_sync}
\end{figure}

\ns{To verify the decentralized time synchronization method externally, and to demonstrate the system performance improvement with time synchronization compared to the same system (frequency syntonized) without continuous compensation of the time biases, multiple experiments were conducted. In the first experiment, the time synchronization method was applied at the start of the experiment and then disabled to test the effect of the dynamic bias on the clock time synchronization. The result in Fig.~\ref{without_time_sync} shows the temporal drift in the time offset between RF front-ends. The result from the second experiment, plotted in Fig.~\ref{with_time_sync}, shows the improvement of the time synchronization of the system resulting from the continuous evaluation and compensation of the dynamic bias in the system, which cannot be compensated for by frequency locking the system.}

\begin{figure}[t!]
\centering
\includegraphics[width=1.0\columnwidth]{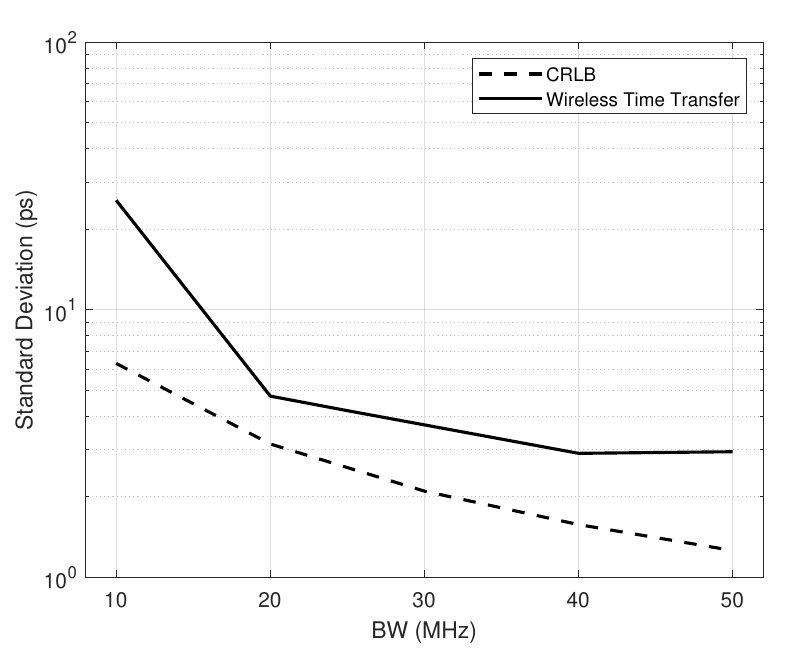}
\caption{Comparison between the standard deviation from the wireless time transfer configuration and the CRLB over a sweep of tone separation (10 MHz to 50 MHz in 10 MHz steps). The link SNRs were uniform and the only parameter changing was the tone separation.}
\label{BW_wireless_p}
\end{figure}

\ns{Fig.~\ref{BW_wireless_p} shows the standard deviation of the wireless time transfer configuration over a sweep of waveform bandwidth (tone separation) from 10 MHz to 50MHz, along with the CRLB (the dashed curve) at fixed SNR. The standard deviation of the time offsets at each tone separation was calculated from ten repeated measurements during each of 100 iterations. The average of the computed standard deviations was calculated, and the worst precision after 40 iterations was recorded and plotted at each 
tone separation as shown in the figure. Note that the calculated precisions from the measured data closely follow the trend of the CRLB. At higher bandwidths approaching the Nyquist rate, discretization errors became more prominent, leading to a leveling or increase in the standard deviation.}

\begin{figure}[t!]
\centering
\includegraphics[width=1.0\columnwidth]{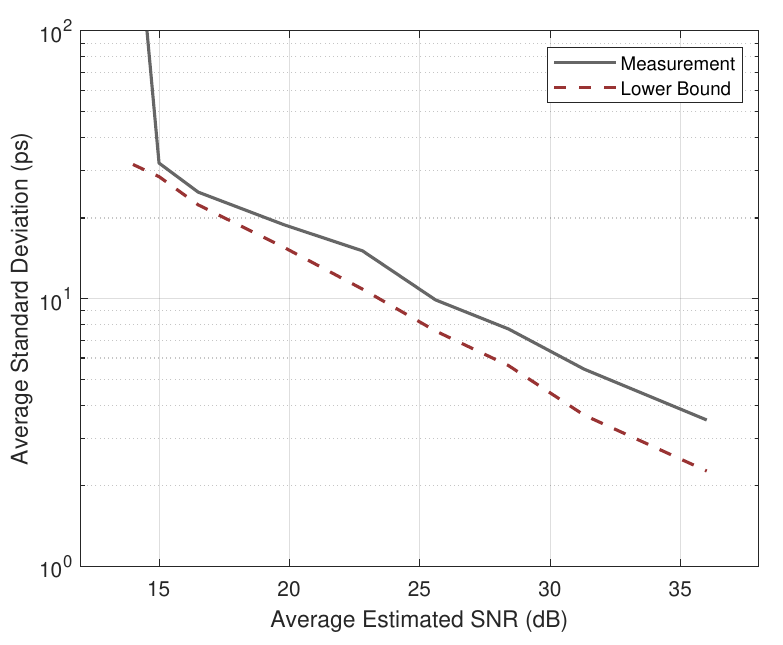}
\caption{\ns{Comparison between the standard deviation of the wireless time transfer configuration and the lower bound \eqref{lb}. The variances of the individual link in the four-node array were evaluated based on the estimated SNRs for each link. The standard deviation of the network was computed from the average variance. The variance of the lower bound was estimated separately for each link using the estimated SNR and averaged. The lower bound is the square root of the average variance. The results are plotted over a range of the average estimated SNRs for all the links in the network from 14 dB to 36 dB.}}
\label{SNR_wireless_p}
\end{figure}

\ns{Finally, the standard deviation from the averaged variances of all the time offsets was computed over a sweep of the average estimated SNR and plotted in Fig.~\ref{SNR_wireless_p} for the wireless time transfer configuration. The lower limit derived in \eqref{lb} is included for comparison and performance evaluation. This figure shows that as the average SNR increased, the average precision increased following the lower bound closely. It is important to note that the lower bound defined in \eqref{var} describes the time delay estimation lower bound for one channel. In Fig.~\ref{BW_wireless_p} the focus is on evaluating the precision of the system with respect to the tone separation. Therefore, the average standard deviation of the time offsets between the nodes in the network was evaluated under the assumption that the channel SNRs were uniform. For a multi-node array, the average CRLB in a network with uniform channel SNRs equals the CRLB for one channel. Practically, variation of network channel SNRs over time and across the channels is due to many factors; among these are differing relative distances between the nodes, which affect the signal power, and differences in the channel noise levels.}

\section{Conclusions}

In this paper, a novel decentralized time-alignment method for a distributed antenna array using the average consensus algorithm and two-way time transfer was implemented and validated. The method achieved better than 3 ps standard deviation at 36 dB SNR and 40 MHz tone separation for a wireless time transfer among the four nodes of a distributed array. The clock time for all the nodes in the system reached convergence within 20 iterations to within 12 ps total error in a decentralized approach. For a distributed antenna array, this supports distributed coherent transmission of communications signals with bandwidths of up to 8.33 GHz with a gain of 90\% of the ideal beamforming gain at a probability of 90\%~\cite{nanzer2017open,8378649}.

\bibliographystyle{IEEEtran}
\bibliography{IEEEabrv, DTAM_1}
\end{document}